\documentclass[twocolumn,showpacs,preprintnumbers,amsmath,amssymb, superscriptaddress, prl]{revtex4}

\usepackage{graphicx}
\usepackage{dcolumn}
\usepackage{bm}
\usepackage{color}
\definecolor{red}{rgb}{1,0,0}

\definecolor{green}{rgb}{0,1,0}

\definecolor{blue}{rgb}{0,0,1}

\begin{document}
\preprint{APS}

\title{Persistent spin dynamics intrinsic to amplitude-modulated long-range magnetic order}

\author{M. Pregelj}
\affiliation{Jo\v{z}ef Stefan Institute, Jamova cesta 39, 1000 Ljubljana, Slovenia}
\affiliation{Laboratory for Neutron Scattering, PSI, CH-5232 Villigen, Switzerland}
\author{A. Zorko}
\affiliation{Jo\v{z}ef Stefan Institute, Jamova cesta 39, 1000 Ljubljana, Slovenia}
\affiliation{EN-FIST Centre of Excellence, Dunajska 156, SI-1000 Ljubljana, Slovenia}
\author{O. Zaharko}
\affiliation{Laboratory for Neutron Scattering, PSI, CH-5232 Villigen, Switzerland}
\author{D. Ar\v{c}on}
\affiliation{Jo\v{z}ef Stefan Institute, Jamova cesta 39, 1000 Ljubljana, Slovenia}
\affiliation{Faculty of Mathematics and Physics, University of Ljubljana, Jadranska cesta 19, SI-1000 Ljubljana, Slovenia}
\author{M. Komelj}
\affiliation{Jo\v{z}ef Stefan Institute, Jamova cesta 39, 1000 Ljubljana, Slovenia}
\author{A. D. Hillier}
\affiliation{ISIS Facility, Rutherford Appleton Laboratory, Chilton, Didcot, Oxon OX11 OQX, United Kingdom}
\author{H. Berger}
\affiliation{\'{E}cole Polytechnique F\'{e}d\'{e}rale de Lausanne,  Lausanne, Switzerland}
\date{\today}

\begin{abstract}

An incommensurate elliptical helical magnetic structure in the frustrated coupled-spin-chain system FeTe$_2$O$_5$Br is surprisingly found to persist down to 53(3)\,mK ($T/T_N$\,$\sim$\,1/200), according to neutron scattering and muon spin relaxation.
In this state, finite spin fluctuations at $T$\,$\to$\,0 are evidenced by muon depolarization, which is in agreement with specific-heat data indicating the presence of both gapless and gapped excitations. 
We thus show that the amplitude-modulated magnetic order intrinsically accommodates contradictory persistent spin dynamics and long-range order and can serve as a model structure to investigate their coexistence.
\end{abstract}

\maketitle

Geometrical frustration is a common precursor for exotic magnetic ground states -- from long-range ordered (LRO) incommensurate (IC) spiral states to highly disordered frozen spin-glass states \cite{Lacroix}. It can even lead to spin liquids, where spin fluctuations may endure down to zero temperature -- persistent spin dynamics (PSD) \cite{Uemura94} is present.
Such behavior is typically found in highly-frustrated pyrochlore and kagom\'{e} spin systems with macroscopically degenerate ground states \cite{Balents}.
Since spin fluctuations hinder the onset of extended static correlations, PSD and LRO are generally considered mutually exclusive.
Remarkably, their coexistence in the same phase has been reported in several frustrated magnetic systems \cite{Yaouanc, Lago, Tb2Sn2O7, Giblin, McClarty, Zheng, Dunsiger, Rule}, but still lacks a suitable explanation.

\begin{figure}[b]
\includegraphics[width=85mm,keepaspectratio=true,angle=0,trim=2mm 10mm 10mm 10mm]{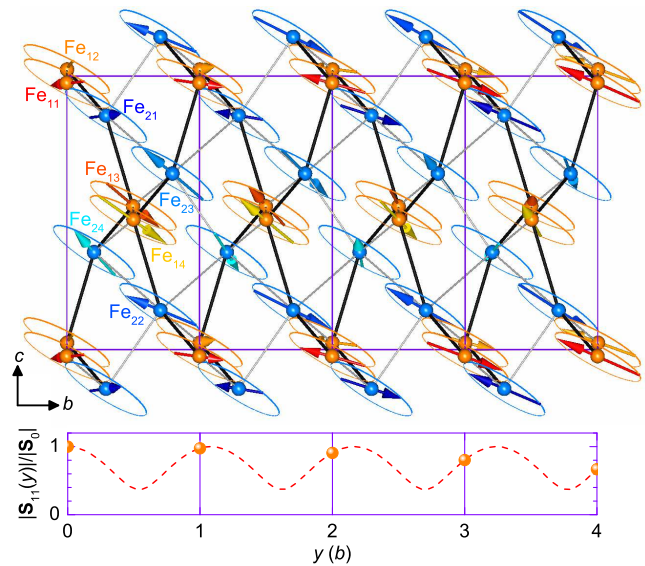}
\caption {(color online) Top: Magnetic structure in FeTe$_2$O$_5$Br at 2\,K. Light (orange) and dark (blue) spheres are Fe$_1$ and Fe$_2$ magnetic ions, respectively. The shades (colors) of the magnetic moments (arrows) are different for the eight magnetically inequivalent sites Fe$_{mn}$. The amplitude of the moment changes with its orientation within the ellipsoidal envelope. Black and gray lines denote the dominant intrachain and the weaker interchain exchange interactions, respectively. Bottom:  IC amplitude modulation of the Fe$_{11}$ magnetic moment along the $b$-axis.}
\label{mag-struc}
\end{figure}

To explore this phenomenon we focus on the FeTe$_2$O$_5$Br multiferroic, in which the magnetic exchange network consists of alternating  Fe$^{3+}$ ($S$\,=\,5/2) spin chains coupled by weaker frustrated interactions within the $bc$ layers \cite{PregeljAFMR} (Fig.\,\ref{mag-struc}).
The magnetic order at 5\,K, i.e., well below the N\'{e}el temperature $T_N$\,=\,10.5\,K, was described as an IC collinear amplitude-modulated (AMOD) structure, with magnetic vector $\bf{q}$\,=\,($\frac{1}{2}$\,0.463\;0) \cite{Pregelj1}. 
In such a state only part of the total Fe magnetic moment at each site contributes to LRO, while its counterpart (at each site) is disordered and may fluctuate.
Generally, on cooling the ordered component fully develops, which manifests either as a ''squaring'' of the AMOD structure \cite{squaring} or as a consecutive transition, where a perpendicular ordered component develops leading to a circular helix \cite{Lawes}. Since FeTe$_2$O$_5$Br exhibits no subsequent transition down to 300\,mK \cite{Pregelj2}, it is possible that PSD and LRO coexist.

In this letter, we report on a combined study of spherical neutron polarimetry (SNP) and neutron diffraction, which reveals that the IC AMOD magnetic structure persists to the lowest accessible temperatures ($T/T_N$\,$\sim$\,1/175). 
This is consistent with muon spin relaxation ($\mu$SR) measurements at $T/T_N$\,$\sim$\,1/200, which in addition to static LRO signify the presence of PSD.
The coexistence of LRO and PSD is supported by specific-heat data, indicating gapped as well as gapless magnetic excitations.
Our study suggests that this is intrinsic to AMOD magnetic structures. It offers a well-defined framework and a coherent explanation for the coexistence of LRO and PSD that has been missing for the known cases \cite{McClarty}.

We begin by a reinvestigation of the FeTe$_2$O$_5$Br magnetic ground state, combining single-crystal neutron diffraction and SNP, which was proven before to be invaluable for determination of complex magnetic structures \cite{NeutronScatt}. The experiments were performed at 5 and 2\,K, respectively, on high-quality single crystals \cite{Pregelj1} with an average size of $15\times8\times2$\,mm$^3$ at the Swiss Neutron Spallation Source (SINQ), Paul Scherrer Institute (PSI), Switzerland. For SNP the MuPAD device on the triple axis spectrometer TASP ($\lambda$\,=\,3.2\,\AA) was used, while the diffraction experiment employed the single-crystal diffractometer TriCS ($\lambda$\,=\,2.317\,\AA).
Polarization matrices were measured for three different crystal orientations \cite{supplementary}, accessing for the first time \cite{Pregelj1, Zaharko} also $hkl$\,$\pm$\,$\bf{q}$ (l\,$\neq$\,0) magnetic reflections.

\begin{figure}[!]
\includegraphics[width=85mm,keepaspectratio=true,angle=0,trim=0mm 5mm 0mm 5mm]{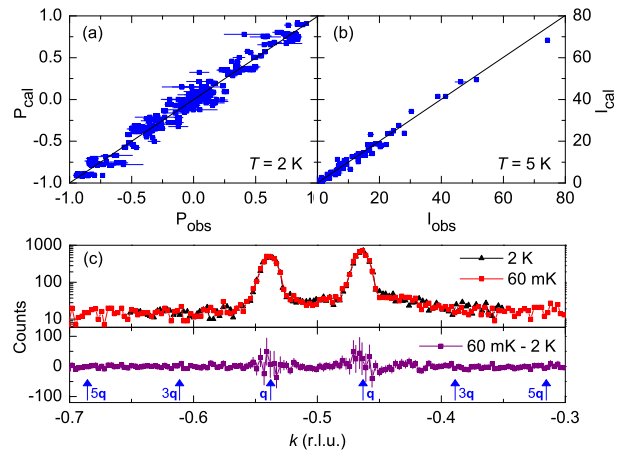}
\caption {Refinement quality for (a) polarization matrices and (b) integrated intensities. (c) Top: $k$-scans at 60(3)\,mK and 2\,K across (3.5~-0.5$\pm\delta$\,0) reflections, witnessing the absence of higher harmonics, expected to occur at marked positions. Bottom: The difference between the two scans which match within the error bar.}
\label{neutrons}
\end{figure}

\begin{table} [!]
\caption{Parameters of the best magnetic structure model at 2\,K for two independent magnetic atoms (Fe$_1$ and Fe$_2$), and eight magnetic phases $\psi_{mn}$, i.e., one for each of the magnetic Fe$_{mn}$ atoms in the unit cell ($m$\,=\,1,2, $n$\,=\,1-4). The sites Fe$_{12}$-Fe$_{14}$ are obtained from Fe$_{11}$ [$0.1184(6)$, $-0.001(1)$, $0.9734(7)$] and Fe$_{22}$-Fe$_{24}$ from Fe$_{21}$ [$0.9377(6)$, $0.2953(1)$, $0.8562(6)$] by symmetry elements $i$, $2_{1y}$ and $2_{1y}i$, respectively. The orientation of the moments is given in the $a^*bc$ coordinate system. 
\label{tab111}}
\begin{ruledtabular}
\begin{tabular}{c|c|c|c|c||c|c|c}
     $s$ = Re, Im   & Fe$_1^{\text{Re}}$ & Fe$_1^{\text{Im}}$ & Fe$_2^{\text{Re}}$ & Fe$_2^{\text{Im}}$ & $n$ &  $\psi_{1n}$ & $\psi_{2n}$  \\
\hline
$S_{0\,x}^s$ /$|S_{0}^s|$        & 0.70   & 0.71   &   0.64 &   0.76 & 1 &   0.00 &  0.00 \\
$S_{0\,y}^s$ /$|S_{0}^s|$        & 0.70   &-0.67   &   0.76 &  -0.61 & 2 &  0.04 &  0.93 \\
$S_{0\,z}^s$ /$|S_{0}^s|$        &-0.14   & 0.20   &  -0.10 &   0.21 & 3 &    0.17 &  0.20 \\ \cline{1-5}
$|{\bf S}_{0\,m}^s|/|{\bf S}_{0}|$  & 0.39   & 0.96   &  0.35  &  1.00  & 4 &    0.21 &  0.15 \\
\end{tabular}
\end{ruledtabular}
\end{table}
The general magnetic structure model, in which the magnetic order breaks all crystallographic symmetry operations \cite{Pregelj1, Zaharko}, dictates the magnetic moment at a particular Fe site to follow an elliptical helix with the pitch along the magnetic $\bf{q}$ vector,
\small 
\begin{equation}
{\bf{S}}_{mn}({\bf{r}}_{i})  =  {\bf{S}}_{0\,mn}^{\text{Re}}\cos({\bf{q}}\cdot{\bf{r}}_{i} - \psi_{mn}) 
 +  {\bf{S}}_{0\,mn}^{\text{Im}}\sin( {\bf{q}}\cdot{\bf{r}}_{i} - \psi_{mn}). 
\end{equation} 
\normalsize
Here, vector ${\bf{r}}_i$ defines the origin of the $i$-th cell, $m$\,=\,1,2 identifies the crystallographically inequivalent Fe-sites, and $n$=1-4 denotes the four Fe positions within the crystallographic unit cell (see caption of Table\,\ref{tab111}). The complex vector ${\bf{S}}_{0\,mn}$ is determined by its real and imaginary components, ${\bf{S}}_{0\,mn}^{\text{Re}}$ and ${\bf{S}}_{0\,mn}^{\text{Im}}$, defining the amplitude and the orientation of the magnetic moments, while $\psi_{mn}$ denotes a phase shift.
We assume the same moment ${\bf{S}}_{0\,mn}$\,$\equiv$\,${\bf{S}}_{0\,m}$ for all crystallographically equivalent Fe-sites.
The reliability of the refined magnetic structure is ensured by simultaneous refinement of all SNP and integrated magnetic-peak intensities data \cite{supplementary}. 
The best solution (Fig.\,\ref{mag-struc} and Table\,\ref{tab111}) with the dominant component pointing $\sim$\,45\,$^\circ$ away from $a$ towards --$b$ axis agrees with the previously proposed simplified collinear IC AMOD structure \cite{Pregelj1}. In addition, a small perpendicular component is identified, resulting in an overall elongated-elliptical cycloid ($|{\bf{S}}_0^{\text{Re}}|/|{\bf{S}}_0^{\text{Im}}|$\,$\sim$\,0.37), with its normal canted $\sim$\,15\,$^\circ$ away from the $c$-axis. The new data thus reveal a magnetic structure that combines AMOD and helical properties.
The quality of the new refinement reflects in the reduced $\chi^2$\,=\,9.6 of the polarization matrices being notably decreased with respect to its value in the collinear AMOD (14.7) and the circular helical structures (14.9) \cite{supplementary}.
A small misfit [Figs.\,\ref{neutrons}(a) and (b)] inevitably originates from a weak nuclear-magnetic-interference term \cite{supplementary}.

At $T$\,$\to$\,0, the ordered component of the magnetic moments in the elliptical IC AMOD structure is expected to grow at the expense of the disordered one. 
This can reflect either as a change of the magnetic-reflection intensities, if circular helix is formed, or alternatively, as additional magnetic reflections with propagation vectors 3$\bf{q}$, 5$\bf{q}$,~... and intensities 1/9, 1/25,~... of the first-order magnetic reflections in case of ''squaring'' \cite{squaring}.
To detect such changes, single crystal neutron diffraction was performed at 60(3)\,mK ($T/T_N$\,$\sim$\,1/175), where we recorded broad $k$-scans of several $hkl$\,$\pm$\,$\bf{q}$ pairs, i.e., intersecting the positions of $hkl$\,$\pm$\,$n\bf{q}$, $n$\,=\,3,\,5,\,~... reflections.  
Surprisingly, we find no significant difference between 60\,mK and 2\,K data as well as 
no trace of higher harmonics [Fig.\,\ref{neutrons}(c)], which implies that both the AMOD ordered component and its disordered counterpart are still present at the lowest accessible temperatures.

\begin{figure}[!]
\includegraphics[width=85mm,keepaspectratio=true,angle=0,trim=0mm 5mm 0mm 10mm]{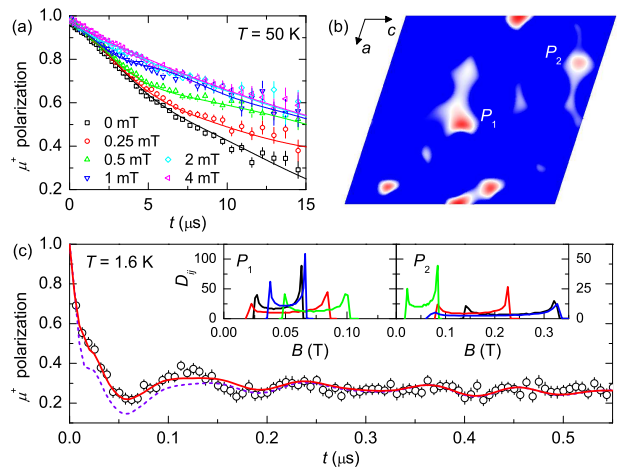}
\caption {(a) Longitudinal-field $\mu$SR measurements on powder sample FeTe$_2$O$_5$Br at 50\,K (symbols) and corresponding fits (lines) -- see text. (b) Electrostatic potential at $b$\,=\,0.15, i.e., intersecting the most pronounced minima, including $P_1$ and $P_2$. (c) Zero-field $\mu$SR measurement at 1.6\,K with corresponding two- (dashed line) and three-component (solid line) fits given by the magnetic structure (see text). Insets: calculated field distributions $D_{ij}(B)$ at $P_i$ ($i$\,=\,1,2) sites for the four magnetically inequivalent positions $j$ within the unit cell.}
\label{muSR-HT}
\end{figure}

Since the neutron diffraction experiments probe only static magnetism, we employed a local-probe $\mu$SR technique, which is extremely sensitive to internal magnetic fields and can distinguish between fluctuating and static magnetism, as well as between LRO and static magnetic disorder \cite{muScience}.
The $\mu$SR-experiments were performed on the MUSR instrument at the ISIS facility, Rutherford Appleton Laboratory, United Kingdom, on the same high-purity powder samples as used in our earlier study \cite{supplementary, ZorkoMuSR}. All the data in the following are shown with properly subtracted background signal ($\sim$\,15\,\%) \cite{supplementary}.
Preliminary results indicated that muons stop at several inequivalent positions \cite{ZorkoMuSR}.
However, to test the magnetic structure model, these must be precisely determined.
New measurements were therefore first performed in the paramagnetic state, at 50\,K [Fig.\,\ref{muSR-HT}(a)], where weak static nuclear magnetic fields are expected to govern the $\mu^+$ spin relaxation \cite{muScience}. Since these fields can be exactly calculated from the crystal structure, they are essential for identification of the muon stopping sites, as demonstrated below.
In the case of a single muon stopping site, in a paramagnetic powder sample the muon polarization in zero applied magnetic field (ZF) is given by the Gaussian Kubo-Toyabe relaxation function $G_{\text{KT}}(t,\Delta)$\,=\,$\tfrac{1}{3}+\tfrac{2}{3}[1-(\Delta t)^2]\exp[-(\Delta t)^2/2]$ multiplied by the exponential function $\exp[-\lambda_L t]$. 
The former accounts for a static Gaussian nuclear field distribution with the width $\Delta/\gamma_\mu$ ($\gamma_\mu$\,=\,2$\pi\times$135.5\,MHz/T), whereas the latter describes a weak dynamical relaxation due to fast electronic fluctuations \cite{muScience}. The resulting function has a single dip, which is removed only by a fast dynamical relaxation, i.e., $\lambda_L$\,$\gtrsim$\,$\Delta$.
Our ZF data exhibits a more complex behavior [Fig.\,\ref{muSR-HT}(a)], as they require two two-component model
\begin{equation}
\label{eq:muSR50K}
G(t)=[G_{\text{KT}}(t,\Delta_A)+G_{\text{KT}}(t,\Delta_B)]\cdot\exp[-\lambda_L t]. 
\end{equation}
This model is supported by measurements in longitudinal applied magnetic fields (LF), where decoupling of the muon relaxation from nuclear-magnetic fields occurs in two steps at $\sim$\,0.5 and $\sim$\,2\,mT [Fig.\,\ref{muSR-HT}(a)], respectively. 
Simultaneous fit of all 50\,K $\mu$SR data to the two-component model, extended for the case of LF \cite{muScience}  [Fig.\,\ref{muSR-HT}(a)], yields the relative occupancies  of the two muon stopping sites of 83(3)\,\% and 17(3)\,\% with corresponding $\Delta_A/\gamma_\mu$\,=\,0.064(5)\,mT and $\Delta_B/\gamma_\mu$\,=\,0.304(5)\,mT, whereas $\lambda_L$\,=\,0.039(3)\,$\mu$s$^{-1}$. 

Since $\mu^+$ are positively charged particles and are generally expected to stop at the electrostatic-potential minima, we calculated the electrostatic potential in the FeTe$_2$O$_5$Br unit cell using density functional theory \cite{supplementary}. 
This way several possible stopping sites were identified [Fig.\,\ref{muSR-HT}(b)], for which dipolar nuclear-magnetic field distributions were calculated \cite{supplementary}. The dominant $\Delta_A/\gamma_\mu$ is found to be in excellent agreement with nuclear field distributions calculated at two local electrostatic-potential minima, $P_1$\,=\,(0.59,\,0.12,\,0.43) and $P_2$\,=\,(0.25,\,0.15,\,0.90) [Fig.\,\ref{muSR-HT}(b)]. These are thus assigned as the prime muon stopping sites.
On the contrary, $\Delta_B/\gamma_\mu$ does not agree with calculated distribution at any electrostatic-potential-minima \cite{supplementary}, so the third (least occupied) muon stopping site remains unassigned.

Identification of the $P_1$ and $P_2$ stopping sites allows us to calculate local dipolar magnetic fields from the ordered Fe moments and thus to double-check the magnetic structure determined by neutrons.
We computed \cite{supplementary} normalized field distributions $D_{ij}(B)$ at four inequivalent sites ($j$\,=\,1-4)  [Wyckoff position 4($e$)] for each $P_i$ ($i$=1,2) felt by muons stopping in random unit cells [insets in Fig.\,\ref{muSR-HT}(c)].
In Fig.\,\ref{muSR-HT}(c), the ZF $\mu$SR data collected at 1.6\,K at PSI \cite{ZorkoMuSR} is shown together with the fit (dashed line) to the corresponding model 
\begin{equation} \label{eq:muSR}
\begin{split}
G(t)=\tfrac{1}{3}\exp[-(\lambda_L t)^\alpha]+\tfrac{2}{3}\sum_{i,j}[A_i\exp(-\lambda_{i} t) \\
\times\int_0^\infty D_{ij}(B)\cos(\gamma_\mu B t)\mathrm{d}B].
\end{split}
\end{equation}
In this expression the first term, commonly called the "1/3-tail" \cite{muScience}, describes muons in a powder sample, whose initial polarization is parallel to the internal magnetic field and therefore changes only due to fluctuations of this field.
This term (dynamical relaxation rate $\lambda_L$  and stretch exponent $\alpha$) is determined from long-time decay measurements presented in the next paragraph. 
Now we focus on the second term, which depicts oscillations due to internal fields induced by the LRO magnetic order.
These oscillations are damped ($\lambda_i$) by spin fluctuations and/or static relaxation resulting from a distribution of muon stopping positions around $P_i$ \cite{supplementary}, with occupancy $A_i$ ($\sum_iA_i$\,=\,1).
The fit yields $\lambda_1$\,=\,0.60(5)\,$\mu$s$^{-1}$, $\lambda_2$\,=\,50(8)\,$\mu$s$^{-1}$ and $A_2$/$A_1$\,=\,1.4(1). 
Most importantly, our model excellently accounts for the experimental oscillation frequencies of the muon polarization determined by $D_{ij}(B)$ (no adjustable parameters) and thus firmly affirms the elongated-elliptical IC AMOD state.
The small discrepancy [Fig.\,\ref{muSR-HT}(c)] is most likely due to the neglected 17(3)\,\% of muons, with unknown stopping position ($\Delta_B$). Indeed, adding a third component with $A_3$\,=\,17\,\% and its decay approximated by $\exp(-\lambda_3 t)$ leads to a perfect agreement with the experiment [Fig.\,\ref{muSR-HT}(c)].
The improved model yields $\lambda_3$\,=\,11(1)\,$\mu$s$^{-1}$, while parameters $\lambda_1$, $\lambda_2$ and $A_2/A_1$ stay within the error bars of the two-component model.
The agreement of the neutron and the $\mu$SR results proves that $\mu$SR probes the intrinsic magnetic properties and can thus provide an invaluable insight into spin dynamics of the FeTe$_2$O$_5$Br system.

\begin{figure}[!]
\includegraphics[width=85mm,keepaspectratio=true,angle=0,trim=0mm 5mm 0mm 10mm]{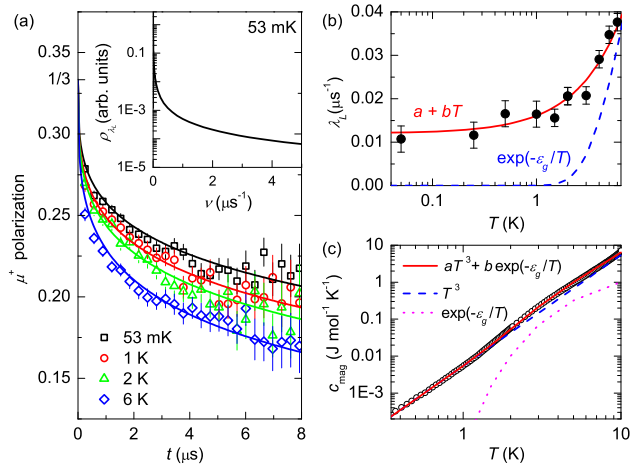}
\caption {(a) Muon depolarization (symbols) at low temperatures, indicating the presence of PSD, and corresponding fits (lines) to the stretched-exponential function $\exp[-(\lambda_L\,t)^{0.3}]$. Inset: relaxation rate distribution at 53\,mK determined from fit (see text). (b) Extracted relaxation rates, showing linear temperature dependence (solid line). For comparison activation behavior (dashed line) for gapped ($\epsilon_g$\,=\,11.5\,K) magnon excitations is shown. (c) Temperature dependence of the magnetic contribution to the specific heat, $c_{\text{mag}}(T)$, together with corresponding fits (see text for details).}
\label{muSR-LT}
\end{figure}

In case of completely static ($\lambda_L$\,=\,0) local magnetic fields, the "1/3 tail" persists in the $t$\,$\to$\,$\infty$ limit [Eq.\,\eqref{eq:muSR}].
Since preliminary ZF $\mu$SR data implied its decay \cite{ZorkoMuSR}, we extended these measurements to longer times and very low temperatures [Fig.\,\ref{muSR-LT}(a)].
The new data confirm the decay of the "1/3 tail" and clearly show its persistence down to the lowest accessible temperature of 53(3)\,mK, i.e., $T/T_N$\,$\sim$\,1/200. Since this decay can only be of dynamical origin, it unambiguously proves that muons, experiencing static magnetic fields due to the IC AMOD order, experience also local-field fluctuations, i.e., revealing the coexistence of PSD and LRO at $T\to0$.

To obtain deeper insight into PSD, we focus on the dynamic part of the $\mu$SR signal, i.e., muon polarization for $t$\,$>$\,0.5\,$\mu$s, where the second term in Eq.\,\eqref{eq:muSR} is already relaxed [Fig.\,\ref{muSR-HT}(b)].
These data are fitted with $G(t)$\,=\,$\tfrac{1}{3}\exp[-(\lambda_L t)^\alpha]$ [Fig.\,\ref{muSR-LT}(a)], where $\alpha$ accounts for the dynamical relaxation rate distribution $\rho_{\lambda_L}(\nu)$, related to the stretched-exponential function by the Laplace transform $\exp[-(\lambda_L t)^\alpha]$\,=\,$\int_0^\infty \rho_{\lambda_L}(\nu)\exp[-(\nu t)]d\nu$ \cite{stretch}.
A small $\alpha$\,=\,0.30(2), found temperature independent below $T_N$, indicates a very broad $\rho_{\lambda_L}(\nu)$ [inset in Fig.\,\ref{muSR-LT}(a)], which most likely reflects 
a distribution and dynamical nature of the disordered parts of the magnetic moments in the AMOD state.
Additionally, $\rho_{\lambda_L}(\nu)$ can be broadened because three different muon stopping sites are present. 
The obtained $\lambda_L$ [Fig.\,\ref{muSR-LT}(b)] shows a linear temperature dependence and, most importantly, converges to a finite zero-temperature value $\lambda_L^0$\,=\,0.012(2)\,$\mu$s$^{-1}$, characteristic of PSD.  
We note that in spite of the small $\lambda_L^0$, the broad $\rho_{\lambda_L}(\nu)$ spans far into the $\mu$SR time window ($\nu$\,$>$0.1\,$\mu$s$^{-1}$), as evident from substantial experimental decay of the ''1/3-tail'' at early times [Fig.\,\ref{muSR-LT}(a)].

In LRO antiferromagnets, $\lambda_L(T)$ is expected to follow either $T^n$ dependence with $n$\,$>$\,2 for $T$\,$\gg$\,$\epsilon_g$ or exp($-\epsilon_g/T$) dependence for $T$\,$\ll$\,$\epsilon_g$ \cite{McClarty, Yoshida, Moriya, Mila}. Here $\epsilon_g$ is the magnon energy gap.
Considering $\epsilon_g$\,=\,11.5\,K, as determined in the recent antiferromagnetic-resonance study \cite{PregeljAFMR}, $\lambda_L(T)$ should change exponentially with temperature in the inspected temperature range, which is clearly not the case [Fig.\,\ref{muSR-LT}(b)].
Hence, $\lambda_L(T)$ signifies magnetic excitations, which are different from the usual magnon modes in LRO states.
Finite relaxation at $T$\,$\to$\,0 is typically found in gapless spin-liquids where it is ascribed to quantum spin fluctuations \cite{Balents, Uemura94, Gardner, Keren}. In FeTe$_2$O$_5$Br, these most probably originate from the disordered component of the magnetic moment at each Fe site, which naturally accompanies the ordered component in an IC AMOD structure.
The dual nature of the ground state is confirmed by the magnetic contribution $c_{mag}(T)$ to specific-heat \cite{Pregelj2}, which is proportional to $T^3$ below $T_N$ and exhibits an additional broad hump around 3\,K [Fig.\,\ref{muSR-LT}(c)]. The former implies three-dimensional antiferromagnetic gapless excitations \cite{Lago, Quilliam, Ohkoshi, DeJongh}, while the latter reveals additional thermally activated exp($-\epsilon_g/T$) term due to the gapped magnetic excitations \cite{Quilliam}.

Finally, we point out similar spin dynamics in the magnetic ground state of volborthite \cite{Yoshida}, where spin-density-wave-like modulation \cite{Yoshida} and IC spin correlations \cite{Nilsen} were found.
Moreover, recent calculations \cite{JansonO} showed that volborthite, like FeTe$_2$O$_5$Br \cite{PregeljAFMR}, has to be treated as a frustrated coupled-spin-chain system. This suggests that such systems are keen to form IC AMOD magnetic ground state, which offers a sound phenomenological explanation of the coexistence of PSD and LRO.

In conclusion, we have found that in FeTe$_2$O$_5$Br the elongated-elliptical IC AMOD magnetic structure persists down to $T/T_N$\,$\sim$\,1/200, as a result of frustrated chain topology.
In this LRO state fluctuations of the remaining disordered spin component at each magnetic site are intrinsic and they manifest as PSD at $T$\,$\to$\,0.
Similar observations in volborthite suggest that IC AMOD magnetic order is a natural habitat for PSD and can serve as a model structure inherently encompassing the intriguing coexistence of PSD and LRO.
This conjecture could be tested on linarite, for which frustrated chains were recently reported to induce a LRO AMOD structure \cite{linarite}. Further in-depth theoretical investigations are required to account for PSD in IC AMOD structures on the microscopic level.

\acknowledgments
We thank M. J. P. Gingras for valuable comments. 
This research project has been partially supported by the Slovenian Research Agency, Project No. J1-2118,  by the Swiss National Science Foundation Project No. 200021-129899, and by the European Commission under the 7th Framework Program through the 'Research Infrastructures' action of the 'Capacities' Program, NMI3-II Grant number 283883. Contract No: CP-CSA\_INFRA-2008-1.1.1 Number 226507-NMI3.


\begin{widetext}
\vspace{15cm}
\begin{center}

{\large {\bf Supplementary material:\\
 Persistent spin dynamics intrinsic to amplitude-modulated long-range magnetic order}}\\
\vspace{0.3cm}
M. Pregelj$^{1,2}$, A. Zorko$^{1,3}$, O. Zaharko$^2$, D. Ar\v{c}on$^{1,4}$, M. Komelj$^1$, A. D. Hillier$^5$, H. Berger$^6$

\vspace{0.2cm}

{\small {\it
$^1$Jo\v{z}ef Stefan Institute, Jamova cesta 39, 1000 Ljubljana, Slovenia\\
$^2$Laboratory for Neutron Scattering, PSI, CH-5232 Villigen, Switzerland\\
$^3$EN-FIST Centre of Excellence, Dunajska 156, SI-1000 Ljubljana, Slovenia\\
$^4$Faculty of mathematics and physics, University of Ljubljana, Jadranska cesta 19, SI-1000 Ljubljana, Slovenia\\
$^5$ISIS Facility, Rutherford Appleton Laboratory, Chilton, Didcot, Oxon OX11 OQX, United Kingdom\\
$^6$\'{E}cole Polytechnique F\'{e}d\'{e}rale de Lausanne,  Lausanne, Switzerland\\
}}

\end{center}
\end{widetext}

\section{Neutron experiments and magnetic structure refinement}
Spherical neutron polarimetry experiment was performed for three different crystal orientations. In addition to the $hk0$ orientation (O1), where the scattering plane was defined by the (1\,0\,0) and (0\,1\,0) reciprocal vectors \cite{Pregelj1}, the crystal was rotated to the scattering plane defined by the (0\,1\,0) and (1\,0\,2) vectors, (orientation\,2 - O2). Two additional orientations (O3a) and (O3b) corresponded to the scattering plane with the (0\,0\,1) and either $\bf{q}_a$\,=\,(0.5\,0.463\,0) or $\bf{q}_b$\,=\,(0.5\,0.537\,0) vectors. This was achieved by tilting the crystal away from (1\,1\,0) by $\pm$1.7\,$^\circ$.
The 54 accumulated polarization matrices were obtained for 45 different magnetic reflections, i.e., some of them were measured in two different crystal orientations or with reversed incoming neutron polarization.

To estimate the experimental uncertainty and to eliminate its influence on the modeling, we proceeded as follows. First, we took into account leakage of the opposite polarization in the incoming and scattered beams, by considering  imperfect, 97.5(5) $\%$, efficiency of the instruments polarizers.
Second, since synchrotron data \cite{Pregelj1, PregeljAFMR} showed that the change of the crystal structure at $T_N$ is very subtle and is not yet fully determined, we had to neglect the existence of the nuclear-magnetic interference term and to assign purely magnetic origin to all IC reflections. To avoid unphysical results, we therefore increased the estimated standard deviations of the matrix elements so that they within the error bar obey the general symmetry of the polarization matrix for the purely magnetic case, e.g., $P_{yy}= -P_{zz}$, $P_{yx}=P_{zx}$, ... \cite{Blume, Maleev, Brown1, NeutronScatt}.

Applying the above approximations we performed several simulated annealing runs, refining all datasets (including 87 integrated magnetic-peak intensities from the conventional neutron diffraction experiment) to the same magnetic structure model, but at the same time allowing different domain populations for each experiment, i.e., separately for O1, O2, O3 and for integrated intensities \cite{MagOpt}.
The minimized total cost function is defined as cost$_{\text{tot}}$\,=$\,\sum_j \chi_j^2$/($N_{j\,\text{obs}}$--$N_{\text{par}}$).
Here $j$ is the number of the datasets and $N_{\text{par}}$ is the number of the fitting parameters.
For each dataset with $N_{{j\,\text{obs}}}$ observations we define $\chi_j^2$\,=\,$\sum_{i=1}^{N_{j\,\text{obs}}}(X_{i\,\text{obs}}$\,--\,$X_{i\,\text{calc}})^2/\sigma_{X_{i\,\text{obs}}}^2$,  where $X$ denotes polarization matrix elements $P$ and/or integrated intensities $I$, and $\sigma_{X_{i\,\text{obs}}}$ is the estimated standard deviation of the observation.
For the best magnetic structure model (Table\,1 in the main text) the resulting total cost is cost$_{\text{tot}}$\,=\,33.7 and partial costs cost$_{\text{P}}$=$\chi_P^2$/($N_{P
\text{obs}}$--$N_{\text{par}}$)\,=\,9.6 for polarization matrices and cost$_{\text{I}}$=$\chi_I^2$/($N_{I
\text{obs}}$--$N_{\text{par}}$)\,=\,24.1 for integrated intensity. This is significantly better than cost$_{\text{tot}}$\,=\,36.1, obtained for the collinear AMOD model with cost$_{\text{P}}$\,=\,14.7 and cost$_{\text{I}}$\,=\,21.4 and that of the cycloidal (full moment) model cost$_{\text{tot}}$\,=\,36.2 with cost$_{\text{P}}$\,=\,14.9 and cost$_{\text{I}}$\,=\,21.3, especially when considering that polarization matrices hold more detailed directional information.
\begin{figure}[!]
\includegraphics[width=85mm,keepaspectratio=true,angle=0,trim=0mm 20mm 0mm 10mm]{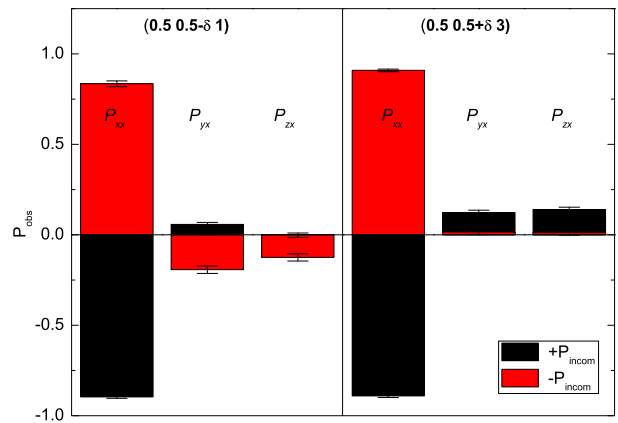}
\caption {P$_{ix}$ ($i$\,=\,$x,y,z$) elements for ($h$ $k\pm\delta$ $l$) reflections for + and -- incoming polarizations, signifying the effect of the nuclear magnetic interference.}
\label{SNP-nuc}
\end{figure}
A small deviation between the calculated and the observed polarization matrix elements is largely due to a weak nuclear-magnetic-interference term, which is most pronounced in $P_{ix}$, $i$\,=\,$y,z$ matrix elements, where the first and the second indexes indicate the polarization of the incoming and outgoing neutron polarizations, respectively. 
In particular, we where able to model the change of the $P_{ix}$ elements on reversal of
the incident polarization (Fig.\,\ref{SNP-nuc}) only when considering additional nuclear-magnetic-interference term -- the magnetic chiral contribution alone was not sufficient \cite{Blume, Maleev, Brown1, NeutronScatt}. 
The apparent nuclear-magnetic interference suggests the presence of higher order ($>$\,2) terms in the magnetoelectric coupling that may explain a small induced electric polarization \cite{Pregelj1}.

\section{$\mu$SR data analysis}
\begin{figure}[!]
\includegraphics[width=85mm,keepaspectratio=true,angle=0,trim=0mm 5mm 0mm 10mm]{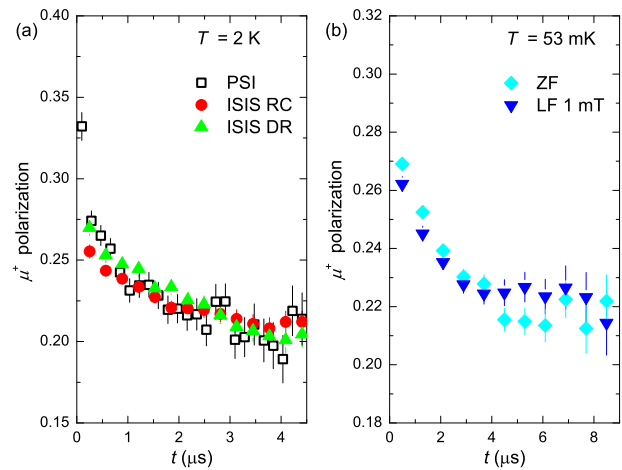}
\caption {(a) Comparison of the ZF $\mu$SR data for 2\,K obtained at different instruments and experimental setups after background subtraction. (b) Comparison of muon depolarization in ZF and in longitudinal applied field (LF) of 1\,mT measured at ISIS.}
\label{supp-muSR}
\end{figure}

The use of powder samples is essential to ensure that exactly 1/3 of muons have initial polarization parallel to a local magnetic field, which is most important for further analysis. To ensure the quality of the powder samples several well-characterized single crystals were grinded.
We took special care of the background signal, whose presence is inevitable in the MUSR ISIS (MUSR instrument at the ISIS facility, Rutherford Appleton Laboratory, United Kingdom) setups due to muons stopping in the sample holder, cryostat tail, etc.
As we combined data from two ISIS setups [a dilution refrigerator (DR) and a regular cryostat (RC)] as well as previous data from PSI, all with different backgrounds, these need to be properly subtracted from the data.
The value of the background for each setup is set by two constraints: (i) from a non-frozen fraction at low-temperature [amplitude of the remaining oscillating part in a weak transverse-field (TF) experiment below $T_N$ -- contribution of muons that are not exposed to the static electronic magnetic field from the LRO and precess around the applied TF field], and (ii) by matching the ZF depolarization curves measured at the same temperature.
We got about 15\,\% (depending on the setup) of background, which is typical for the used ISIS instrument (with given slit opening and sample size) and matches well with the known relaxation of a silver sample holder. Additionally, the subtracted ISIS and PSI data, all taken at $T$\,=\,2\,K, match well with each other [Fig.\,\ref{supp-muSR}(a)].

Nuclear-magnetic field distributions at electrostatic-potential minima (Table\,\ref{tabDFT}) were calculated  by randomly orienting all nuclear magnetic moments (with appropriate abundance) within a fictitious Lorentz sphere with radius large enough to ensure convergence. 

Similar approach was used for calculating the electronic magnetic field distributions at $P_1$ and $P_2$ sites in the LRO magnetic state. These distributions were exactly determined by magnetic structure model and acquire finite widths due to IC modulation of the magnetic structure along the $b$-axis.
The (contact) hyperfine interaction between muons and electrons is neglected. There are several justifications for this. First, both known muon stopping sites $P_1$ and $P_2$ lie well outside the exchange paths (outside $bc$ planes defining the plane of coupled chains). The distance of both to the nearest iron is more than 2.5\,\AA, whereas the nearest oxygen is 1.35\,\AA \, away (Table\,\ref{tabDFT}). Therefore, we do not expect significant electron density on these sites. The latter distance is in reasonable agreement with the usual distance $\sim$\,1.0\,\AA, expected when muon bond to oxygen \cite{muScience}. Second, the calculated dipolar fields match perfectly with observed oscillations in muon polarization (with no adjustable parameters).
We stress that the above agreement indicates that muons probe intrinsic magnetic properties of the  elongated-elliptical IC AMOD magnetic structure and do not distort their local surrounding.
In addition, high-resolution synchrotron data and low-temperature magnetic susceptibility do not show any sign of crystalline disorder or magnetic impurities, respectively \cite{Pregelj1, Pregelj2}. Therefore, the observed dynamics cannot originate from such effects.

In search for the possible third muon stopping site, we calculated nuclear-magnetic field distributions throughout the unit cell. This allowed us to identify positions with fields corresponding to experimentally determined value. Since these positions are far from the local electrostatic-potential minima, we speculate that the corresponding muons may deform their local environment and change the electrostatic potential.
Given the high field strength at this sites, their positions have to be away from stopping sites $P_1$ and $P_2$ in the vicinity of oxygens. One of the possibilities is that they are close to Br atoms, which possess the highest nuclear magnetic moments.

Finally, field decoupling at lowest temperatures implies that relaxation of the ''1/3 tail'' is not due to nuclear field relaxation on the unknown stopping site [$\Delta_B/\gamma_\mu$\,=\,0.304(5) mT] and thus show that it is indeed dynamical [Fig.\,\ref{supp-muSR}(b)].

\section{Density-functional-theory calculations}
The electrostatic potential was calculated {\it ab initio} within 
the framework of the density-functional theory and the local-density
approximation (LDA) \cite{Perdew:1992-1} of the exchange-correlation
potential. We applied the Wien97 code \cite{Blaha:1990}, which adopts the 
full-potential linearized-augmented-plane-waves (FLAPW) 
method \cite{Wimmer:1981}. The calculations were performed for the structure
with the experimental atomic positions and the lattice parameters, whereas 
the muffin-tin radia were 2.1\,${\rm a.u.}$
for the Fe atoms, 1.98\,${\rm a.u.}$ for the Te atoms, 1.5\,${\rm a.u.}$
for the O atoms, and 2.68\,${\rm a.u.}$ for the Br atoms. 
The plane-wave-expansion  cut-off energy was set to 16\,Ry, 
and the summation of 333 {\bf k}-vectors from the full Brillouin-zone (BZ) 
was carried out by means of the Gaussian method \cite{Fu:1983} with
the smearing parameter of 0.02\,Ry. 

\begin{table} [!]
\caption{The most pronounced local electrostatic-potential minima $P_i$, with corresponding positions, distances to the closest oxygen ($d_{\text{O}-P_i}$) and iron ($d_{\text{Fe}-P_i}$) atom, and local nuclear-field distribution widths ($\Delta/\gamma_\mu$).
\label{tabDFT}}
\begin{ruledtabular}
\begin{tabular}{c|c c c|c|c|c}
$i$ & $x$ & $y$ & $z$   & $d_{\text{O}-P_i}$(\AA)  & $d_{\text{Fe}-P_i}$(\AA) &  $\Delta/\gamma_\mu$(mT) \\
\hline
1 & 0.59 & 0.13 & 0.43 & 1.36 & 4.50 & 0.071\\
2 & 0.25 & 0.15 & 0.90 & 1.34 & 2.55 & 0.065\\
3 & 1.00 & 0.13 & 0.54 & 1.50 & 2.88 & 0.017\\
4 & 0.91 & 0.15 & 0.66 & 1.51 & 2.88 & 0.018\\
5 & 0.91 & 0.09 & 0.20 & 1.42 & 2.28 & 0.039\\
\end{tabular}
\end{ruledtabular}
\end{table}
The potential was calculated on a discrete mesh containing $100\times100\times100$ points within the unit cell.
From the calculated data we determined the most pronounced local electrostatic-potential minima
given in Table\,\ref{tabDFT}. These were all found in vicinity of oxygen sites, as expected for oxides \cite{muScience}. We note that potential wells around the listed minima significantly differ among each other, i.e., some being very sharp while others very broad [Fig.\,3(b) in the main text]. Latter thus allow muons more freedom when choosing their actual stopping site.

\section{Specific heat - Determination of the lattice contribution}
\begin{figure}[!]
\includegraphics[width=85mm,keepaspectratio=true,angle=0,trim=0mm 5mm 0mm 5mm]{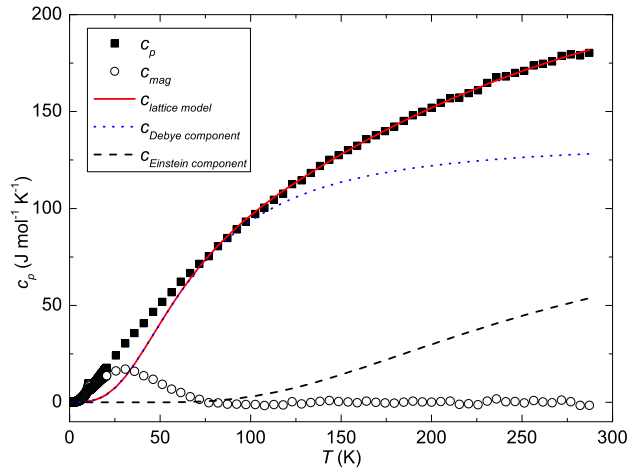}
\caption {Specific heat measurements (solid squares), model of phonon/lattice contribution (solid line), and derived magnetic contribution (open circles). Additional dotted and dashed lines indicate Debye and Einstein contributions, respectively.}
\label{cp-lattice}
\end{figure}

To deduce magnetic contribution to the specific heat $c_{\text{mag}}(T)$, one needs first to estimate phonon contribution due to crystal lattice vibrations. Since there exists no isostructural nonmagnetic compound for FeTe$_2$O$_5$Br, we need to model the high-temperature data, i.e., above 50\,K, where magnetic correlations should already be negligible \cite{Pregelj2}. 
The most common approach is based on the Debye model \cite{Ashcroft}, where vibrational modes are approximated by elastic vibrations of an isotropic continuous body, with a linear dispersion relation. However, this model alone cannot explain our experiments since total specific heat ($c_p$) at high temperatures shows much stronger temperature dependence. In order to explain high-temperature phonon contribution we refer here to far-infrared investigation of the isostructural magnetic compound FeTe$_2$O$_5$Cl \cite{FTOC}, where a number of vibrational modes were identified.
The fact that most of these modes are sharp suggests that they correspond to well-defined nondispersive optical modes. Therefore, in addition to the Debye contribution accounting for (linearly) dispersive acoustic vibrational modes, we took into account the strongest four optical modes, i.e., 320, 400, 470, and 650 cm$^{-1}$, and approximate them with corresponding four Einstein contributions \cite{Ashcroft}. 
As a result we obtained a very good description of the high-temperature data (Fig.\,\ref{cp-lattice}) with the Debye temperature $T_D$\,=\,280\,K.
We note that low-temperature ($T<T_N$) part of the specific heat results can be fitted also with the Debye model alone, which would suggest that $c_{\text{mag}}(T<T_N)$ is absent. However, such model yields completely unphysical behavior above ($T>T_N$), making this model unrealistic.

\section{Low temperature experiments}
Neutron diffraction and muon spin relaxation measurements at lowest temperatures, i.e., in the mK region, were performed in a dilution refrigerator.
To ensure thermalization of the sample, in both experiments samples were prepared and mounted according to the standard procedures.
In the case of single-crystal neutron diffraction the crystal was mounted on a Cu plate, and closed in a Cu container. 
In addition, He was used as exchange gas, which at the temperatures of the experiment became a superfluid film. 
In the $\mu$SR experiment, highly diluted GE varnish was admixed to the FeTe$_2$O$_5$Br powder sample to ensure thermal contact between sample grains and the Au sample holder.
Finally, we stress that in both experiments the samples were held at base temperature for several hours before data acquisition.


\begin{thebibliography}{99}

\bibitem{Lacroix} {\em Introduction to Frustrated Magnetism}, edited by C. Lacroix, P. Mendels, and F. Mila (Springer-Verlag, Berlin, 2011).
\bibitem{Uemura94} Y. J. Uemura et al., {\em Phys. Rev. Lett.} {\bf 73}, 3306 (1994).
\bibitem{Balents} L. Balents, {\em Nature} {\bf 464}, 199 (2010).
\bibitem{Yaouanc} A. Yaouanc et al., {\em Phys. Rev. Lett.} {\bf 95}, 047203 (2005).
\bibitem{Lago} L. Lago et al., {\em J. Phys.: Condens. Matter} {\bf 17}, 979 (2005).
\bibitem{Zheng} X. G. Zheng et al., {\em Phys. Rev. Lett.} {\bf 95}, 057201 (2005).
\bibitem{Tb2Sn2O7} P. Dalmas de R\'{e}otier et al., {\em Phys. Rev. Lett.} {\bf 96}, 127202 (2006)
\bibitem{Dunsiger} S. R. Dunsiger et al., {\em Phys. Rev.} {\bf B 73}, 172418 (2006).
\bibitem{Giblin} S.R. Giblin et al., {\em Phys. Rev. Lett.} {\bf 101}, 237201 (2008).
\bibitem{Rule} K. C. Rule et al., {\em J. Phys.: Condens. Matter} {\bf 21} 486005 (2009).
\bibitem{McClarty} P. A. McClarty et al., {\em J. Phys.: Condens. Matter} {\bf 23} 164216 (2011).
\bibitem{PregeljAFMR} M. Pregelj et al., {\em Phys. Rev.} {\bf B 86}, 054402 (2012).
\bibitem{Pregelj1} M. Pregelj et al., {\em Phys. Rev. Lett.} {\bf 103}, 147202 (2009).
\bibitem{squaring} S.-M. Choi et al., {\em Phys. Rev. Lett.} {\bf 87}, 107001 (2001).
\bibitem{Lawes} G. Lawes et al.,  {\em Phys. Rev. Lett.} {\bf 93}, 247201 (2004).
\bibitem{Pregelj2} M. Pregelj et al., {\em Phys. Rev.} {\bf B 82}, 144438 (2010).
\bibitem{NeutronScatt} {\em Neutron Scattering from magnetic materials}, edited by T. Chatterji (Elsevier B. V., Amsterdam, 2006).
\bibitem{supplementary} see Supplementary Material for details of neutron scattering experiments and magnetic structure refinement, background determination of the $\mu$SR data, nuclear- and electronic-magnetic-field calculations, density-functional-theory calculations, and modeling of the crystal-lattice contribution to the specific heat.
\bibitem{Zaharko} O. Zaharko et al., {\em J. Phys.: Conf. Ser.} \textbf{211}, 012002 (2010).
\bibitem{muScience} A. Youanc and P. Dalmas de R\'{e}otier, {\em Muon spin rotation, relaxation and resonance} (Oxford University Press, Oxford 2011).
\bibitem{ZorkoMuSR} A. Zorko et al.,  {\em J. Appl. Phys.} {\bf 107}, 09D906 (2010).
\bibitem{stretch} D. C. Johnston, {\em  Phys. Rev.} {\bf B 74}, 184430 (2006).
\bibitem{Moriya} T. Moriya {\em Prog. Theor. Phys.} {\bf 16}, 23 (1956).
\bibitem{Mila} F. Mila and T. M. Rice, {\em  Phys. Rev.} {\bf B 40}, 11382 (1989).
\bibitem{Yoshida} M. Yoshida et al. {\em Phys. Rev. Lett.} {\bf 103}, 077207 (2009).
\bibitem{Gardner} J. S. Gardner et al., {\em Phys. Rev. Lett.} {\bf 82}, 1012 (1999).
\bibitem{Keren} A. Keren et al., {\em Phys. Rev. Lett.} {\bf 84}, 3450 (2000).
\bibitem{DeJongh} L. J. De Jongh \& A. R. Miedema, {\em Adv. Phys.}, {\bf 50}, 947 (2001).
\bibitem{Quilliam} J. A. Quilliam et al., {\em Phys. Rev. Lett.} {\bf 99}, 097201 (2007).
\bibitem{Ohkoshi} S. Ohkoshi et al., {\em Coord. Chem. Rev.} {\bf 249}, 1830 (2005).
\bibitem{Nilsen} G. J. Nilsen et al., {\em  Phys. Rev.} {\bf B 84}, 172401 (2011).
\bibitem{JansonO} O. Janson et al., {\em  Phys. Rev.} {\bf B 82}, 104434 (2010).
\bibitem{linarite} B. Willenberg et al., {\em Phys. Rev. Lett.} {\bf 108}, 117202 (2012).



\end{thebibliography}

\begin{thebibliography}{99}


\bibitem{Pregelj1} M. Pregelj et al., {\em Phys. Rev. Lett.} {\bf 103}, 147202 (2009).
\bibitem{PregeljAFMR} M. Pregelj et al., {\em Phys. Rev.} {\bf B 86}, 054402 (2012).
\bibitem{NeutronScatt} {\em Neutron Scattering from magnetic materials}, edited by T. Chatterji (Elsevier B. V., Amsterdam, 2006).


\bibitem{Blume} M. Blume, {\em Phys. Rev.} {\bf 130}, 1670 (1963).
\bibitem{Maleev} S. V. Maleev, V. G. Baryaktar, and R. A. Suris, {\em Sov. Phys. Solid State} {\bf 4}, 2533 (1963).
\bibitem{Brown1} P. J. Brown, {\em Physica} {\bf 297}, 198 (2001).
\bibitem{MagOpt} MagOpt program based on CrysFML library: J. Rodriguez-Carvajal and J. Gonzalez-Platas, {\em Acta Cryst A} {\bf 58}, C87 (2002).


\bibitem{muScience} A. Youanc and P. Dalmas de R\'{e}otier, {\em Muon spin rotation, relaxation and resonance} (Oxford University Press, Oxford 2011).
\bibitem{Pregelj2} M. Pregelj et al., {\em Phys. Rev.} {\bf B 82}, 144438 (2010).

\bibitem{Perdew:1992-1} J.~P.~Perdew and Y.~Wang, {\em Phys. Rev.} {\bf B 45}, 13244 (1992).
\bibitem{Blaha:1990} P.~Blaha, K.~Schwarz and P.~Sorantin and S.~B.~Trickey, {\em Comput. Phys. Commun.} {\bf 59}, 399 (1990).
\bibitem{Wimmer:1981} E.~Wimmer, H.~Krakauer, M.~Weinert and A.~J.~Freeman, {\em Phys. Rev.} {\bf B 24}, 864 (1981).
\bibitem{Fu:1983} C.-L.~Fu and K.-M.~Ho, {\em Phys. Rev.} {\bf B 28}, 5480 (1983).



\bibitem{Ashcroft}  N. W. Ashcroft and N. D. Mermin, {\em Solid State Physics} (Harcourt Brace College Publishers, Fort Worth 1996).
\bibitem{FTOC} F Pfuner et. al., {\em J. Phys.: Condens. Matter} {\bf 21}, 375401 (2009).





\end{thebibliography}
\end{document}